\documentclass[prd,twocolumn,floatfix,nopacs,preprintnumbers,nofootinbib]{revtex4}
\usepackage[utf8x]{inputenc}
\usepackage{mathrsfs}
\usepackage{amssymb}

\usepackage{amsmath}
\usepackage{graphicx}
\usepackage[caption = false]{subfig}
\usepackage[normalem]{ulem}
\usepackage[dvipsnames]{xcolor}
\definecolor{lcolor}{rgb}{0.5,0,0}
\definecolor{citcolor}{rgb}{0,0.3,0.0}
\usepackage{nicefrac}

\usepackage[breaklinks,colorlinks,urlcolor=blue,citecolor=citcolor,linkcolor=lcolor]{hyperref}
\usepackage{mciteplus}
\usepackage{physics}

\newcommand{\lqcd}{\Lambda_{\mathrm{QCD}}}

\newcommand{\jpsi}{$\mathrm{J}/\psi$~}
\newcommand{\jpsim}{\mathrm{J}/\psi}

\newcommand{\mjimwlk}{m_\mathrm{JIMWLK}}

\begin{document}
\begin{abstract}
We perform a global Bayesian analysis of coherent and incoherent diffractive $\mathrm{J}/\psi$ photoproduction in $\gamma+p$ and $\gamma+\mathrm{Pb}$ collisions using a Color Glass Condensate (CGC)-based framework and ultraperipheral collision data from the Large Hadron Collider (LHC), corrected for the expected effect of electromagnetic dissociation (EMD). 
Using Gaussian-process emulators of the underlying CGC calculations, we infer model parameters from a combined set of HERA and LHC measurements. 
We find that the $\gamma+\mathrm{Pb}$ data with EMD correction substantially reduce the previously observed tension between proton and nuclear datasets, enabling a consistent simultaneous description of diffractive $\mathrm{J}/\psi$ production in $\gamma+p$ and $\gamma+\mathrm{Pb}$ collisions within the CGC framework.
\end{abstract}

\author{Heikki Mäntysaari}
\affiliation{Department of Physics, University of Jyväskylä, %
 P.O. Box 35, 40014 University of Jyväskylä, Finland}
\affiliation{Helsinki Institute of Physics, P.O. Box 64, 00014 University of Helsinki, Finland}

\author{Bj\"orn Schenke}
\affiliation{Physics Department, Brookhaven National Laboratory, Upton, New York 11973, USA}

\author{Chun Shen}
\affiliation{Department of Physics and Astronomy, Wayne State University, Detroit, Michigan 48201, USA}

\author{Hendrik Roch}
\affiliation{Department of Physics, University of Jyväskylä, %
 P.O. Box 35, 40014 University of Jyväskylä, Finland}
\affiliation{Helsinki Institute of Physics, P.O. Box 64, 00014 University of Helsinki, Finland}
  
\author{Wenbin Zhao}
\affiliation{Institute of Particle Physics and Key Laboratory of Quark and Lepton Physics (MOE), Central China Normal University, Wuhan, 430079, Hubei, China}

\title{Revisiting the role of saturation in diffractive vector meson production}

\maketitle

\section{Introduction}
\label{sec:introduction}

Understanding the behavior of gluons at small parton momentum fractions $x$ is critical to understand the fundamental properties of hadrons and nuclei. 
At sufficiently small $x$, the rapid growth of gluon density is expected to be moderated by nonlinear recombination effects, leading to the emergence of a saturated state described by the Color Glass Condensate (CGC) effective theory~\cite{Gribov:1983ivg,Mueller:1985wy,McLerran:1993ni,McLerran:1993ka,Iancu:2003xm,Morreale:2021pnn,Garcia-Montero:2025hys} of quantum chromodynamics (QCD). 
Diffractive vector meson production is a sensitive probe of this regime, because at leading order its cross section depends on the square of the target gluon density~\cite{Ryskin:1992ui}. 
It also provides access to the target's spatial structure and structure fluctuations~\cite{Mantysaari:2016ykx,Mantysaari:2020axf}.

CGC-based calculations have successfully described coherent and incoherent diffractive $\mathrm{J}/\psi$ photoproduction in $\gamma+p$ collisions measured at HERA and LHC~\cite{Mantysaari:2016ykx,Mantysaari:2017cni,Mantysaari:2018zdd,Mantysaari:2023qsq,Cepila:2025rkn}, and will play a crucial role in the upcoming Electron Ion Collider (EIC) era~\cite{AbdulKhalek:2021gbh}. 
Applying this framework to ultraperipheral heavy-ion collisions~\cite{Bertulani:2005ru,Klein:2019qfb} enables studies of nuclear gluon distributions, suppression effects, and nuclear structure~\cite{Mantysaari:2026dps}. 
However, calculations constrained by proton data generally predict a stronger energy dependence (less suppression at large photon-nucleus center-of-mass energy $W$) of diffractive $\mathrm{J}/\psi$ production in $\gamma+\mathrm{Pb}$ collisions than observed experimentally, resulting in a persistent tension between proton and nuclear measurements~\cite{Lappi:2013am,Mantysaari:2022sux,Mantysaari:2023xcu,Mantysaari:2024zxq}.
Our recent global Bayesian analysis~\cite{Mantysaari:2025ltq} confirmed this tension. 
We found that a reasonable simultaneous description of $\gamma+p$ and $\gamma+\mathrm{Pb}$ data within the considered framework could not be achieved without introducing an overall $K$ factor in Eq.~\eqref{eq:cross_section} that multiplies all cross sections and is significantly smaller than unity. 
This suppression was compensated by a larger saturation scale, which enhanced the nonlinear dynamics~\cite{Mantysaari:2025ltq}, leading to a stronger suppression of the $\gamma+$Pb cross section at high collision energy. 

Recently, it was demonstrated that electromagnetic dissociation (EMD) can cause exclusive ultraperipheral events to fail experimental exclusivity requirements~\cite{Dyndal:2026uvm}. 
If this occurs, it would lead to a systematic underestimation of the extracted $\gamma+\mathrm{Pb}$ cross sections. 
Correcting for this effect substantially modifies the energy dependence of diffractive $\mathrm{J}/\psi$ production in nuclei.

In this Letter, we revisit the global Bayesian analysis of coherent and incoherent diffractive $\mathrm{J}/\psi$ production in $\gamma+p$ and $\gamma+\mathrm{Pb}$ collisions using experimental data with the additional EMD correction applied~\cite{Dyndal:2026uvm}.
Employing the global analysis framework introduced in our previous work~\cite{Mantysaari:2025ltq}, we perform a new analysis without additional model assumptions. 
We find that the corrected data largely remove the previously observed tension and enable a consistent simultaneous description of proton and nuclear measurements within the CGC framework. 
If the LHC experiments confirm the EMD correction, our result will demonstrate the consistency of gluon saturation effects predicted by the CGC with diffractive vector meson production data off protons and heavy nuclei.

\section{Theoretical framework}
\label{sec:setup}
We calculate coherent and incoherent diffractive \jpsi photoproduction using the same CGC framework as in Refs.~\cite{Mantysaari:2022sux,Mantysaari:2023xcu,Mantysaari:2025ltq,Mantysaari:2025ujz,Mantysaari:2025cok,Mantysaari:2026dps}. The coherent and incoherent cross sections are obtained from the average and variance of the diffractive scattering amplitude $\mathcal{A}$ over target configurations $\Omega$~\cite{Caldwell:2009ke,Kowalski:2006hc,Mantysaari:2020axf},
\begin{align}\label{eq:cross_section}
\begin{split}
\frac{\mathrm d\sigma_{\mathrm{coh}}}{\mathrm dt}
&= \frac{K}{16\pi}
\left|\left\langle \mathcal A \right\rangle_\Omega\right|^2,\\
\frac{\mathrm d\sigma_{\mathrm{inc}}}{\mathrm dt}
&= \frac{K}{16\pi}
\left(
\left\langle |\mathcal A|^2 \right\rangle_\Omega
-
\left|\left\langle \mathcal A \right\rangle_\Omega\right|^2
\right),
\end{split}
\end{align}
where $K$ is an overall normalization factor accounting for uncertainties from the vector meson wave function and higher-order corrections.
It was introduced to test whether the tensions in the model without a $K$ factor could be resolved this way~\cite{Mantysaari:2025ltq}. 

The diffractive scattering amplitude is calculated in the dipole picture as the convolution of the photon--vector meson wave-function overlap with the dipole-target scattering amplitude. 
We employ the boosted Gaussian parametrization for the \jpsi wave function~\cite{Kowalski:2006hc}. 

The dipole amplitude is constructed from Wilson lines using the McLerran-Venugopalan model at $x_0=0.01$ and evolved to smaller $x$ via event-by-event Jalilian-Marian, Iancu, McLerran, Weigert, Leonidov, Kovner (JIMWLK) evolution~\cite{Mueller:2001uk,Lappi:2012vw}.
Proton shape fluctuations are modeled using a constituent hot-spot picture, while nuclear configurations are obtained from Woods-Saxon sampled nucleon positions (with each nucleon constructed using the same geometric model as the proton). 
Further details about the framework can be found in Refs.~\cite{Mantysaari:2025ltq,Mantysaari:2022sux}.

Following Ref.~\cite{Mantysaari:2025ltq}, model parameters are constrained using a multivariate Gaussian likelihood that includes only the diagonal part of the experimental covariance matrix and emulator uncertainties.
Model calculations are replaced by Gaussian-process emulators trained on a large set of full CGC simulations. 
In this work, no new emulator training is required; we directly reuse the validated emulators from Ref.~\cite{Mantysaari:2025ltq} and update the inference analysis using recently published EMD-corrected $\gamma+\mathrm{Pb}$ data. 
Uniform priors are adopted for all model parameters, and posterior distributions are sampled using the \texttt{pocoMC} package~\cite{Karamanis:2022alw,Karamanis:2022ksp}.

The Bayesian analysis constrains the same parameter set as in Ref.~\cite{Mantysaari:2025ltq}. 
The most important parameters characterize the proton geometry and its fluctuations (proton size  $B_G$, hot spot size $B_q$, and magnitude of $Q_s$ fluctuations $\sigma$), the overall color charge density through the ratio $Q_s/(g^2\mu)$, and infrared regulators controlling the long-distance behavior of the Wilson lines and JIMWLK evolution ($m$ and $m_{\mathrm{JIMWLK}}$). 
The evolution speed is governed by the effective scale parameter $\Lambda_{\mathrm{QCD}}$, parametrizing the scale at which the running coupling is evaluated in coordinate space, while an optional overall normalization factor $K$ accounts for residual theoretical uncertainties such as those associated with the vector meson wave function and higher-order corrections. 
Explicit expressions including these parameters are not shown here; we refer the reader to Ref.~\cite{Mantysaari:2025ltq} for details.

The difference relative to Ref.~\cite{Mantysaari:2025ltq} is the replacement of the previously used ultraperipheral collision data by the measurements including the EMD correction. 
This allows us to directly quantify the impact of the EMD-based veto correction of experimental data on the inferred model parameters and on the simultaneous description of $\gamma+p$ and $\gamma+\mathrm{Pb}$ data.

Specifically, we update the dataset used in the Bayesian analysis to include the recently published EMD-corrected integrated $\gamma+\mathrm{Pb}$ cross sections from Ref.~\cite{Dyndal:2026uvm} instead of the original $\gamma+\mathrm{Pb}$ data. 
As mentioned above, these data points lie within the kinematic range covered by the Gaussian-process emulators constructed in Ref.~\cite{Mantysaari:2025ltq}, and no retraining of the emulators is required.

Three integrated cross section data points from the ALICE Collaboration at $W \approx 100\;\mathrm{GeV}$ are not explicitly provided in EMD-corrected form in Ref.~\cite{Dyndal:2026uvm}. 
In this case, we remove the corresponding entries from the emulator-based prediction vector used in the likelihood. 
The EMD-corrected dataset includes an ATLAS point in the same $W$ region.
We include this point in the fit by interpolating the emulator prediction as a function of $W$ within the corresponding kinematical region.

The $|t|$-differential $\gamma+\mathrm{Pb}$ measurements from ALICE~\cite{ALICE:2021tyx,ALICE:2023gcs} are not yet available in EMD-corrected form. 
To avoid introducing a systematic bias from this inconsistency, we do not include their absolute normalization in the fit. 
Instead, we rescale both the coherent and incoherent $|t|$ spectra to match the lowest-$|t|$ points using a single combined scaling factor for the $|t|$-differential emulator predictions for this system. 
This effectively includes only $|t|$-shape information from this dataset in the likelihood. 
This procedure ensures that the nuclear data constrain the $t$-dependence independently of the overall normalization, which is fixed by the integrated EMD-corrected cross sections.

\section{Results}
We perform a new global Bayesian analysis of diffractive \jpsi production in $\gamma+p$ and $\gamma+\mathrm{Pb}$ collisions, including the datasets discussed in Sec.~\ref{sec:setup}.

\begin{figure*}[!tb]
    \centering
    \includegraphics[width=\linewidth]{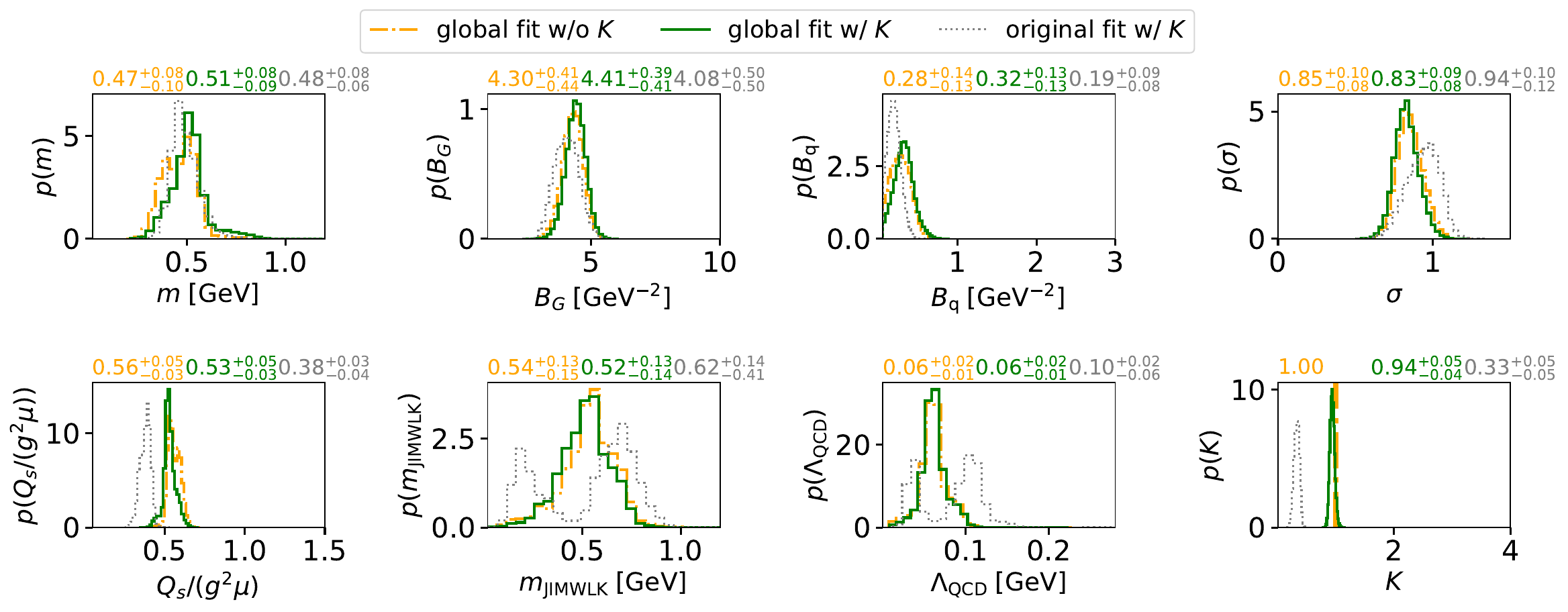}
    \caption{Posterior distributions from the global analysis with (full) and without (dash-dotted) the normalization factor $K$. Dotted curves show results from Ref.~\cite{Mantysaari:2025ltq}, where $K\neq1$ and the bimodal posterior reflects tension between proton and nuclear observables. Numbers above each panel denote the median and 90\% credible interval; shown parameter ranges correspond to the prior ranges.
    }
    \label{fig:corner_full}
\end{figure*}

We first perform a global Bayesian analysis including both datasets. 
To assess whether the additional normalization freedom introduced in Ref.~\cite{Mantysaari:2025ltq} remains necessary when using the nuclear data with EMD correction, we perform fits both with and without the normalization factor $K$.
The corresponding posterior distributions for the combined fits are shown in Fig.~\ref{fig:corner_full}. 
The results with and without the $K$ factor are nearly indistinguishable.

For comparison, we also show the corresponding posterior distribution obtained in Ref.~\cite{Mantysaari:2025ltq} using the setup with the $K$ factor.
That analysis included published experimental data (i.e., no additional EMD correction from~\cite{Dyndal:2026uvm}), and found that the parameters related to the JIMWLK evolution ($\mjimwlk$, $\lqcd$) exhibit multimodality, and furthermore a relatively small $K\sim 0.3$ was required. 
These features were interpreted~\cite{Mantysaari:2025ltq} to originate from the fact that within the applied setup, there is some tension between the $\gamma+p$ and $\gamma+\mathrm{Pb}$ datasets. 
These features are not visible anymore when the data with EMD-correction are used in the fit. 
Instead, both $\mjimwlk$ and $\lqcd$ are well constrained, and the additional normalization freedom is not required by the data anymore ($K\approx 1$ is preferred by the Bayesian fit). We note that this does not mean that there are no model uncertainties (which $K$ was supposed to represent), but merely that the model components, such as the vector meson wave function, are compatible with the data as is.

\begin{figure*}[tb]
  \centering
  \subfloat[$\gamma+p\to\jpsim+p$ cross section as a function of center-of-mass energy compared to ALICE~\cite{ALICE:2014eof,ALICE:2018oyo}, H1~\cite{H1:2005dtp,H1:2013okq}, ZEUS~\cite{ZEUS:2002wfj}, and LHCb~\cite{LHCb:2018rcm} data.]{%
    \includegraphics[width=0.48\textwidth]{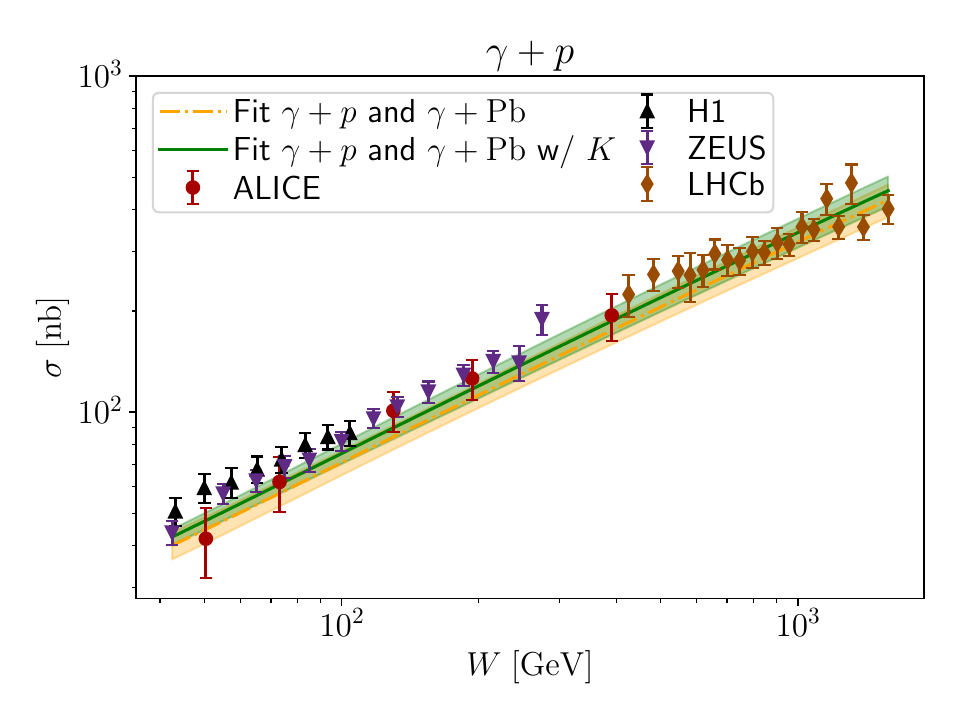}%
    \label{fig:gammap_with_without_K}%
  }
  \hfill
  \subfloat[$\gamma+\mathrm{Pb}\to\jpsim+\mathrm{Pb}$ cross section as a function of center-of-mass energy compared to ALICE~\cite{ALICE:2023jgu} and CMS~\cite{CMS:2023snh} data. The EMD-corrected full points used in the fit are from Ref.~\cite{Dyndal:2026uvm}.]{%
    \includegraphics[width=0.48\textwidth]{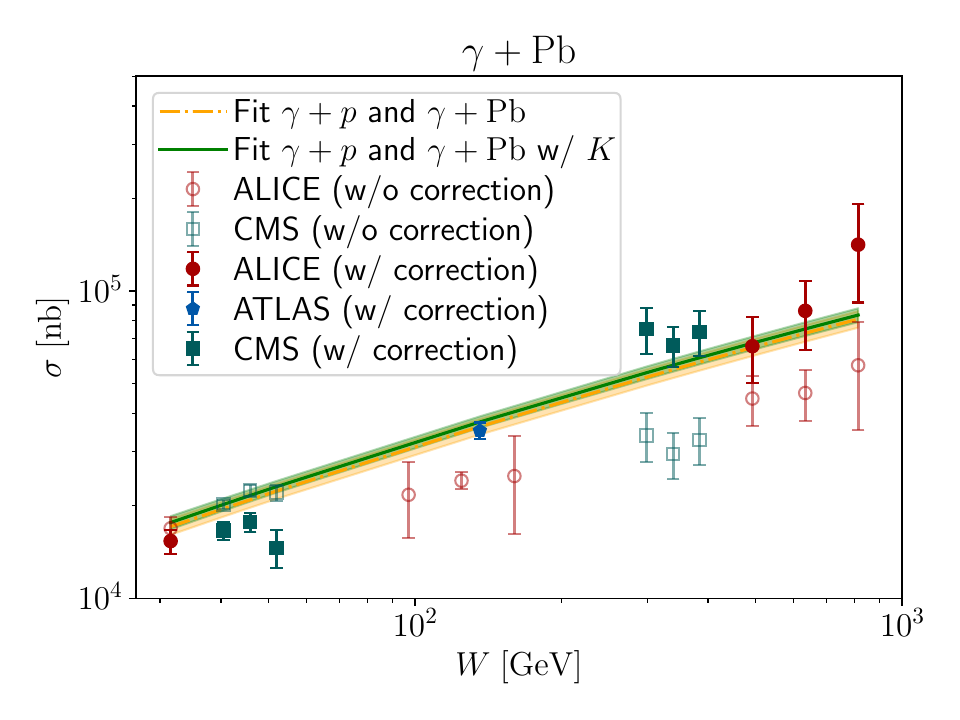}%
    \label{fig:gammaPb_with_without_K}%
  }
  \hfill
  \subfloat[Coherent (smaller $|t|$) and incoherent (larger $|t|$) \jpsi spectra in $\gamma+p$ compared to H1 data~\cite{H1:2013okq}.]{%
    \includegraphics[width=0.48\textwidth]{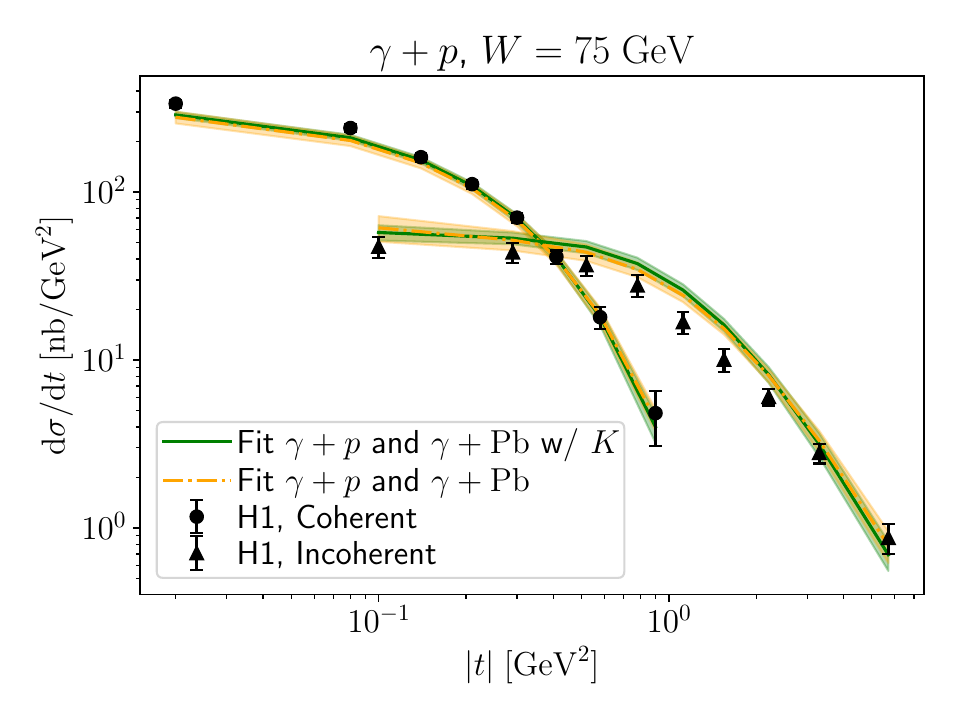}%
    \label{fig:gammap_t}%
    
  }
  \hfill
  \subfloat[Coherent (smaller $|t|$) and incoherent (larger $|t|$) \jpsi spectra in $\gamma+\mathrm{Pb}$ compared to ALICE data~\cite{ALICE:2021tyx,ALICE:2023gcs}. The experimental data is normalized by the $t_{\rm min}$ point and the model prediction by a factor that matches the $t_{\rm min}$ experimental points on average between coherent and incoherent cross sections.]{%
    \includegraphics[width=0.48\textwidth]{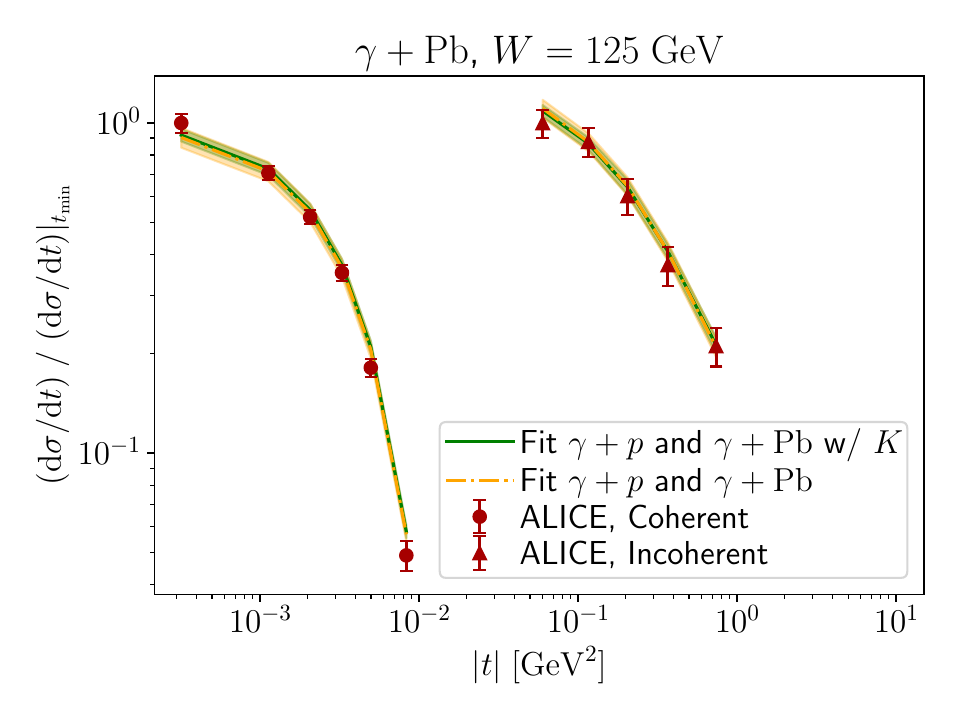}%
    \label{fig:gammaPb_t}%
  }
  \caption{$W$-dependent integrated and $|t|$-differential cross sections from the global fit in the standard parameter setup (dash-dotted) and with the additional $K$ factor (full). Bands indicate 68\% credibility intervals from 25 posterior samples.}
  \label{fig:combined_combinedfits}
\end{figure*}

The corresponding model predictions are shown in Fig.~\ref{fig:combined_combinedfits} with the uncertainty bands reflecting the standard deviation from 25 posterior samples. The combined fit provides a consistent description of both $\gamma+p$ and (EMD-corrected) $\gamma+\mathrm{Pb}$ integrated cross sections.
At the same time, the proton differential spectra are described with the same quality as in Ref.~\cite{Mantysaari:2025ltq}, and the shape of the nuclear $|t|$-distributions is well reproduced. 
Overall, the inclusion of the EMD correction in the data resolves the previously observed tension between proton and nuclear systems within the present CGC framework.
Using the mean and standard deviation from the 25 posterior samples, we obtain reduced chi-square values of $\chi^2_{\rm red}=1.68$ and $1.78$ for the fits with and without the additional $K$ factor, respectively.

The similarity of the posterior distributions and model predictions obtained with and without the $K$ factor suggests that the additional normalization freedom may no longer be required. 
To quantify this observation, we perform a Bayesian model comparison using the Bayes factor, which compares the evidence of competing models after integrating over their parameter spaces~\cite{Gordon:2007xm,Trotta:2008qt}. 
Unlike a comparison based solely on best-fit likelihoods, the Bayes factor penalizes unnecessary model complexity.
To quantify this, we compute Bayes factors comparing the baseline model and the extended model:
\begin{align}
\mathcal B_{A/B} = \frac{\mathcal P(\mathbf y_{\rm exp}|A)}{\mathcal P(\mathbf y_{\rm exp}|B)},    
\end{align}
where $A$ denotes the baseline model with fixed $K=1$, $B$ denotes the extended model with free $K$, and $\mathcal{P}(\mathbf y_{\rm exp}|A)$ denotes the Bayesian evidence of model $A$, obtained by marginalizing the likelihood over the full parameter space.
We find $\ln \mathcal{B}_{A/B} = 2.57 \pm 0.01$, which indicates moderate evidence in favor of the baseline model with fixed $K=1$.
This shows that, despite a slight improvement in the best-fit $\chi^2_{\rm red}$ when introducing the additional normalization parameter, the increase in model complexity is not supported by the data once the full parameter space is taken into account. 
In contrast to the previous analysis, the inclusion of the EMD-correction in the $\gamma+\mathrm{Pb}$ data removes the need for an additional normalization freedom within the present CGC framework. 

\begin{figure*}[!tb]
    \centering
    \includegraphics[width=\linewidth]{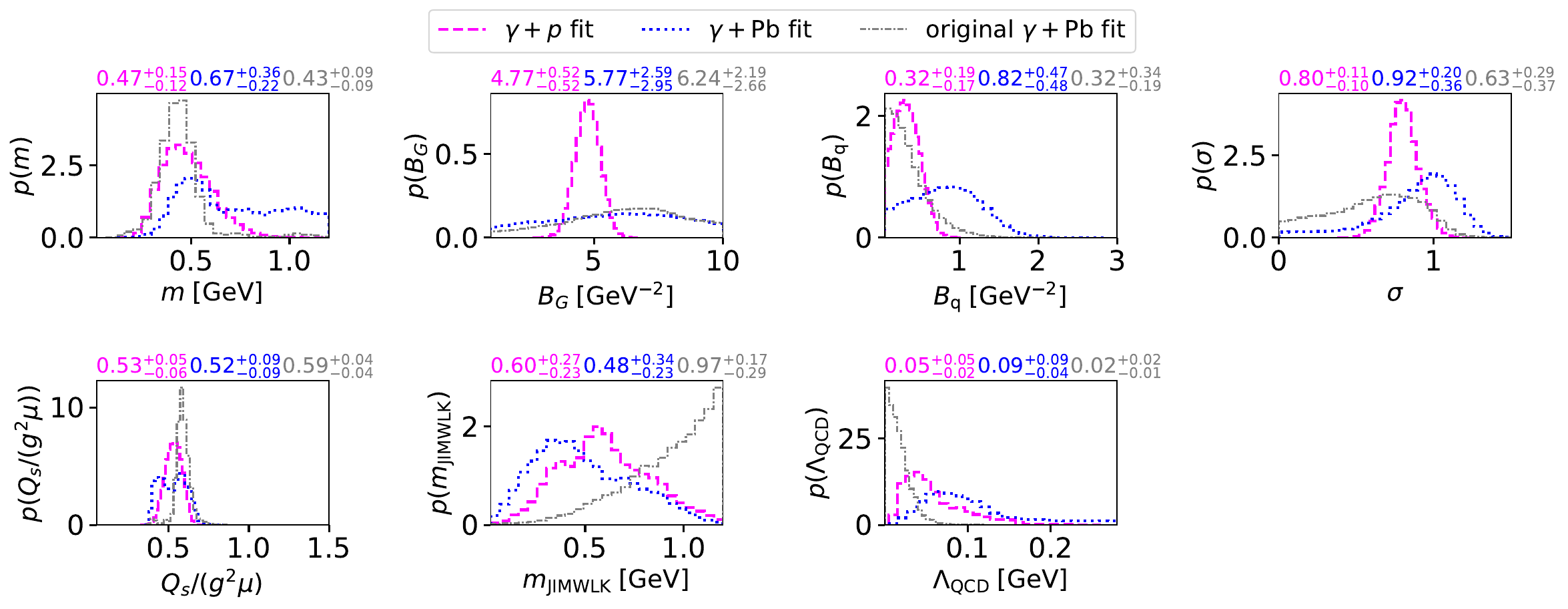}
    \caption{Posterior distributions from separate fits to $\gamma+p$ (dashed) and $\gamma+\mathrm{Pb}$ (dotted) data, compared with the $\gamma+\mathrm{Pb}$ fit from Ref.~\cite{Mantysaari:2025ltq} (dash-dotted). Numbers above each panel give the median and 90\% credible interval; parameter ranges correspond to the priors.}
    \label{fig:corner_eP_ePb_vanilla}
\end{figure*}

To understand why the EMD correction leads to a successful combined fit without requiring an additional normalization factor, we finally perform separate Bayesian fits to the $\gamma+p$ and $\gamma+\mathrm{Pb}$ (with EMD correction) datasets without introducing the overall normalization factor $K$.
The corresponding posterior distributions from the separate fits are shown in Fig.~\ref{fig:corner_eP_ePb_vanilla}.
The proton data constrain the geometric and fluctuation parameters tightly and reproduce the results of Ref.~\cite{Mantysaari:2025ltq}. 
In contrast, the $\gamma+\mathrm{Pb}$ data lead to a noticeably different constraint on the JIMWLK evolution parameters, in particular for $m_{\mathrm{JIMWLK}}$ and $\Lambda_{\mathrm{QCD}}$,  and these parameters now peak at values consistent with the proton fit. 
This differs from the earlier analysis~\cite{Mantysaari:2025ltq}, where the nuclear data preferred values near the boundaries of the prior range. 
Proton geometry parameters remain only weakly constrained by the $\gamma+\mathrm{Pb}$ data.
The shift of the preferred JIMWLK parameters toward values compatible with the proton fit already indicates that a substantial part of the previously observed discrepancy originated from the treatment of the $\gamma+\mathrm{Pb}$ data rather than from a fundamental inconsistency of the CGC framework.

The EMD correction increases the energy dependence of the extracted $\gamma+\mathrm{Pb}$ cross section compared to the previously used measurements. 
As a consequence, the $\gamma+\mathrm{Pb}$ data no longer require the
significantly slower JIMWLK evolution that was preferred in
Ref.~\cite{Mantysaari:2025ltq} and achieved by having a small $K\sim 0.3$ compensated by a larger saturation scale, enhancing nonlinear effects.
This naturally moves the preferred evolution parameters closer to those favored by the proton data.

\begin{figure*}[tb]
  \centering
  \subfloat[$\gamma+p\to\jpsim+p$ cross section as a function of center-of-mass energy compared to ALICE~\cite{ALICE:2014eof,ALICE:2018oyo}, H1~\cite{H1:2005dtp,H1:2013okq}, ZEUS~\cite{ZEUS:2002wfj}, and LHCb~\cite{LHCb:2018rcm} data.]{%
    \includegraphics[width=0.48\textwidth]{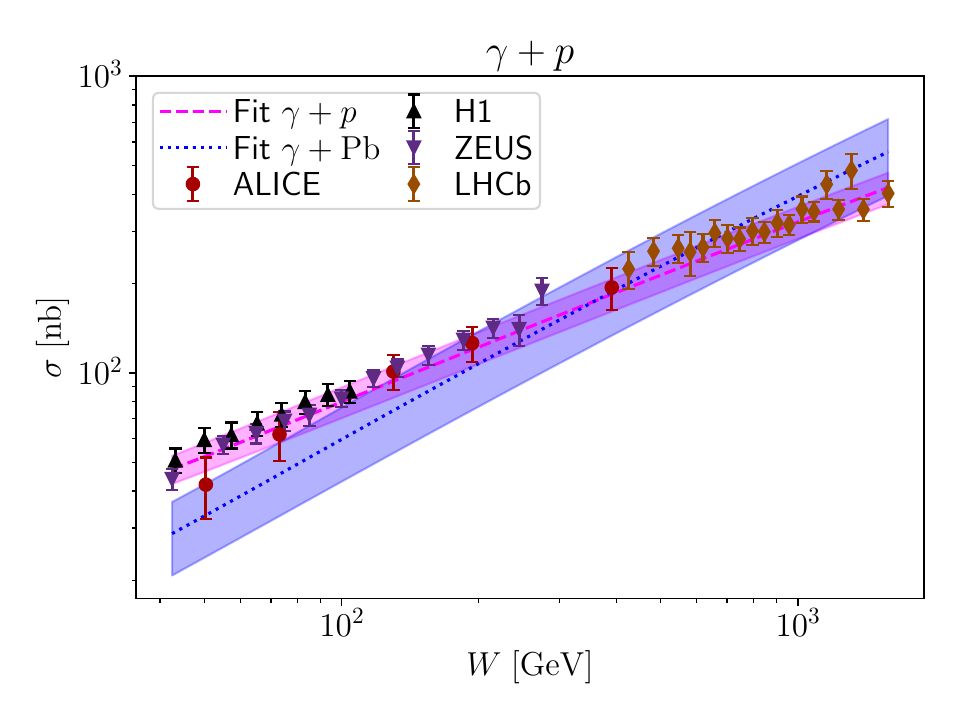}%
    \label{fig:gammap_separatefits}%
  }
  \hfill
  \subfloat[$\gamma+\mathrm{Pb}\to\jpsim+\mathrm{Pb}$ cross section as a function of center-of-mass energy compared to ALICE~\cite{ALICE:2023jgu} and CMS~\cite{CMS:2023snh} data (open circles). The EMD-corrected full points used in the fit are from Ref.~\cite{Dyndal:2026uvm}.]{%
    \includegraphics[width=0.48\textwidth]{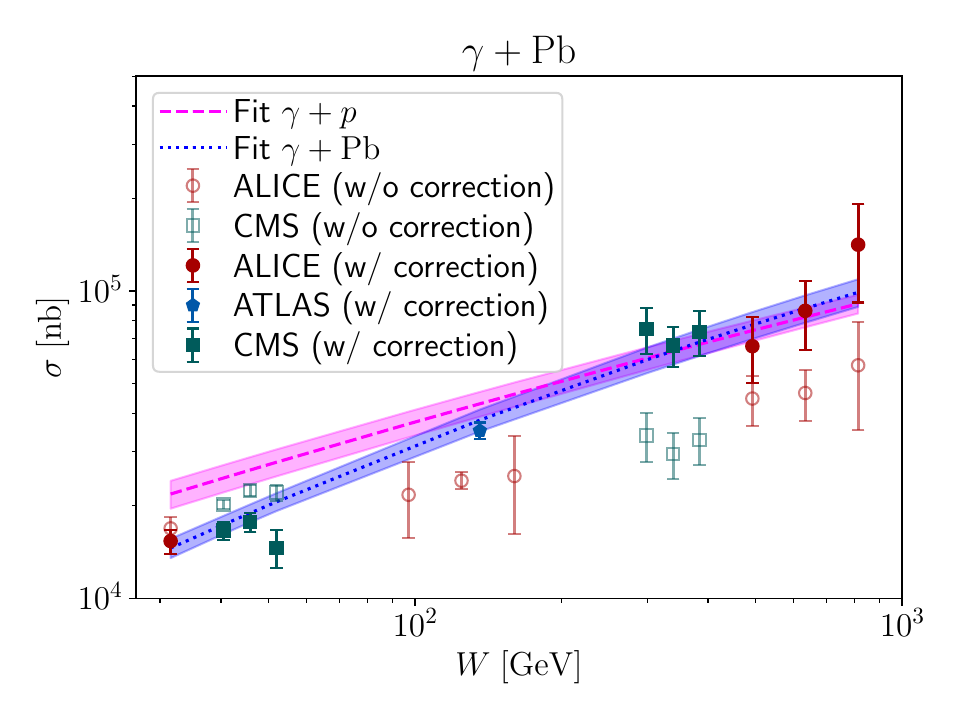}%
    \label{fig:gammaPb_separatefits}%
  }
  \hfill
  \subfloat[Coherent (smaller $|t|$) and incoherent (larger $|t|$) \jpsi spectra in $\gamma+p$ compared to H1 data~\cite{H1:2013okq}.]{%
    \includegraphics[width=0.48\textwidth]{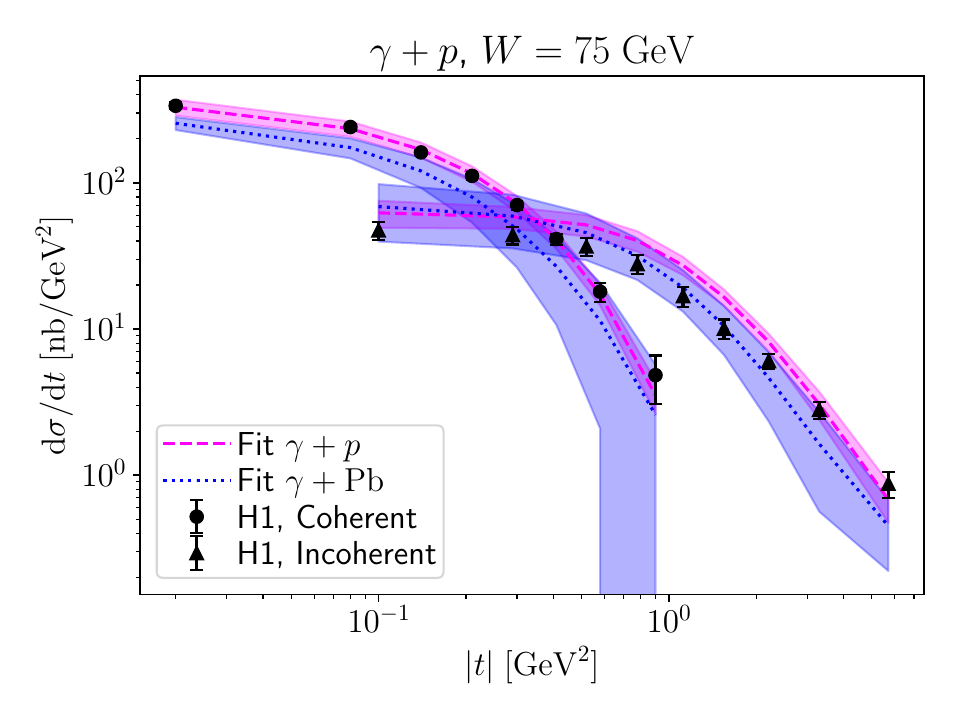}%
    \label{fig:gammap_separatefits_t}%
  }
  \hfill
  \subfloat[Coherent (smaller $|t|$) and incoherent (larger $|t|$) \jpsi spectra in $\gamma+\mathrm{Pb}$ compared to ALICE data~\cite{ALICE:2021tyx,ALICE:2023gcs}.
  The normalization is the same as in Fig.~\ref{fig:gammaPb_t}.]{%
    \includegraphics[width=0.48\textwidth]{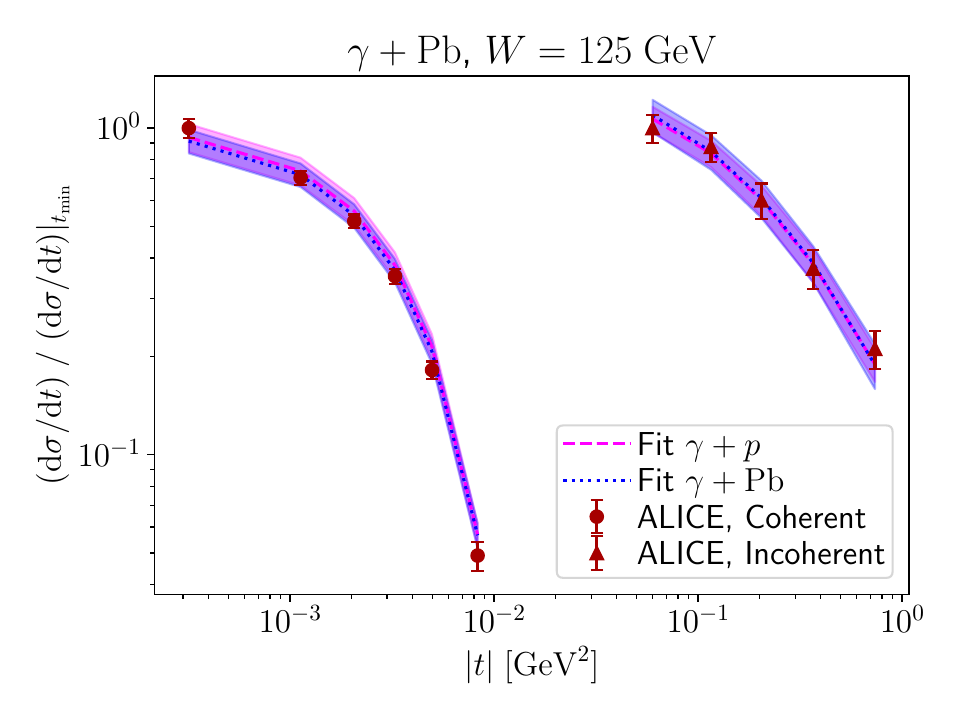}%
    \label{fig:gammaPb_separatefits_t}%
  }
  \caption{Integrated $W$-dependent and $|t|$-differential cross sections from separate fits to $\gamma+p$ (dashed) and $\gamma+\mathrm{Pb}$ (dotted) data using the standard parameter setup (no $K$ factor). Bands indicate 68\% credibility intervals from 25 posterior samples.}
  \label{fig:combined_separatefits}
\end{figure*}

The corresponding predictions for coherent and incoherent cross sections from the separate fits are shown in Fig.~\ref{fig:combined_separatefits}, again with bands indicating the standard deviation from 25 posterior samples.
The $\gamma+p$ fit provides a good description of all proton observables and also reproduces the shape of the $\gamma+\mathrm{Pb}$ $|t|$-differential spectra. 
However, it does not provide a very good description of the energy dependence of the EMD-corrected integrated $\gamma+\mathrm{Pb}$ cross sections.

Conversely, the $\gamma+\mathrm{Pb}$-only fit describes the EMD-corrected nuclear energy dependence and the shape of the $|t|$-spectra in $\gamma+\mathrm{Pb}$ scattering, but leads to a poor description of the $\gamma+p$ data, where the energy dependence becomes too steep. 
Overall, we obtain reduced chi-square values of $\chi^2_{\rm red}=1.16$ and $1.27$ for the fits to only $\gamma+p$ or only $\gamma+\mathrm{Pb}$ data, respectively.

These results are a reflection of the fact that the $\gamma+\mathrm{Pb}$ data alone cannot sufficiently constrain all features of the proton structure (see Fig.~\ref{fig:corner_eP_ePb_vanilla}), while the energy evolution in the dense nuclear system may require additional constraints beyond those provided by the $\gamma+p$ data. These discrepancies do not mean that there is no good combined fit, as we have already shown above. Instead, they mean that fitting using data from one system alone cannot fully constrain the best parameters for both systems.

\section{Conclusions}
We have revisited the global Bayesian analysis of coherent and incoherent diffractive \jpsi production in $\gamma+p$ and $\gamma+\mathrm{Pb}$ collisions including recently published corrections to the ultraperipheral $\mathrm{Pb}+\mathrm{Pb}$ collision data stemming from considering the occurrence of electromagnetic dissociation. 
The EMD-corrected $\gamma+\mathrm{Pb}$ measurements lead to posterior distributions more consistent with proton data and remove the strong preference for an additional normalization $K$ factor observed in our previous global analysis.

Within the present CGC framework, a simultaneous description of proton and nuclear diffractive observables was achieved without additional model freedom.
A Bayes factor analysis found moderate evidence for the model without the additional normalization factor.

We note that the modified data points used here are based on theory-driven corrections~\cite{Dyndal:2026uvm} and have not yet been released as official corrected datasets by the experimental collaborations. 
Further, the $|t|$-differential $\gamma+\mathrm{Pb}$ data constrain the fit only through their shape, as EMD corrections to differential measurements are not yet available. 

Our results demonstrate that electromagnetic-dissociation veto corrections in ultraperipheral collisions can substantially impact the extraction of small-$x$ nuclear dynamics.
We have shown that including the proposed EMD corrections in the $\gamma+\mathrm{Pb}$ data resolves the previously observed tensions between CGC predictions and experiment.

\begin{acknowledgments}
We thank M. Dyndal and L. A. Harland-Lang for providing the EMD-corrected data points.
H.M. and H.R. are supported by the Research Council of Finland, the Centre of Excellence in Quark Matter, and projects 338263 and 359902, and by the European Research Council (ERC, grant agreements No. ERC-2023-101123801 GlueSatLight and ERC-2018-ADG-835105 YoctoLHC). 
This work is supported by the U.S. Department of Energy, Office of Science, Office of Nuclear Physics, under DOE Contract No.~DE-SC0012704 and within the framework of the Saturated Glue (SURGE) Topical Theory Collaboration (B.P.S.).
C.S. is supported by the U.S. Department of Energy, Office of
Science, Office of Nuclear Physics, under DOE Award No. DE-SC0021969 and DE-SCL0000037. C.S. acknowledges a DOE Office of Science Early Career Award.
This research was done using resources provided by the Open Science Grid (OSG)~\cite{Pordes:2007zzb,Sfiligoi:2009cct,OSPool,OSDF}, which is supported by the National Science Foundation awards \#2030508 and \#2323298. 
Part of the numerical simulations presented in this work were performed at the Wayne State Grid, and we gratefully acknowledge their support.
The content of this article does not reflect the official opinion of the European Union and responsibility for the information and views expressed therein lies entirely with the authors.
\end{acknowledgments}

\noindent\textbf{Data availability} Results from the codes~\cite{ipglasma_jimwlk_code,subnucleondiffraction_code,IPGlasmaFramework} are available on Zenodo~\cite{roch_2026_20696856}.


\bibliographystyle{JHEP-2modlong.bst}
\bibliography{refs}

@article{Pordes:2007zzb,
    author = "Pordes, Ruth and others",
    editor = "Keyes, David E.",
    title = "{The Open Science Grid}",
    reportNumber = "FERMILAB-CONF-07-217-CD",
    doi = "10.1088/1742-6596/78/1/012057",
    journal = "J. Phys. Conf. Ser.",
    volume = "78",
    pages = "012057",
    year = "2007"
}

@article{Sfiligoi:2009cct,
    author = "Sfiligoi, Igor and Bradley, Daniel C. and Holzman, Burt and Mhashilkar, Parag and Padhi, Sanjay and Wurthwrin, Frank",
    title = "{The pilot way to Grid resources using glideinWMS}",
    reportNumber = "FERMILAB-CONF-09-373-CD",
    doi = "10.1109/CSIE.2009.950",
    journal = "WRI World Congress",
    volume = "2",
    pages = "428--432",
    year = "2009"
}

@article{Klein:2019qfb,
    author = {Klein, Spencer R. and M\"antysaari, Heikki},
    title = "{Imaging the nucleus with high-energy photons}",
    eprint = "1910.10858",
    archivePrefix = "arXiv",
    primaryClass = "hep-ex",
    doi = "10.1038/s42254-019-0107-6",
    journal = "Nature Rev. Phys.",
    volume = "1",
    number = "11",
    pages = "662--674",
    year = "2019"
}

@Article{Caldwell:2009ke,
     author    = "Caldwell, A. and Kowalski, H.",
     title     = "{Investigating the gluonic structure of nuclei via $\mathrm{J}/\psi$
                  scattering}",
     journal   = "Phys. Rev.",
     volume    = "C81",
     year      = "2010",
     pages     = "025203",
     doi       = "10.1103/PhysRevC.81.025203",
     eprint = "0909.1254",
     archivePrefix = "arXiv",
     SLACcitation  = "%%CITATION = PHRVA,C81,025203;%%"
}

@article{Mantysaari:2016ykx,
    author = {M{\"a}ntysaari, Heikki and Schenke, Bj{\"o}rn},
    title = "{Evidence of strong proton shape fluctuations from incoherent diffraction}",
    eprint = "1603.04349",
    archivePrefix = "arXiv",
    primaryClass = "hep-ph",
    doi = "10.1103/PhysRevLett.117.052301",
    journal = "Phys. Rev. Lett.",
    volume = "117",
    number = "5",
    pages = "052301",
    year = "2016"
}

@article{AbdulKhalek:2021gbh,
    author = "Abdul Khalek, R. and others",
    title = "{Science Requirements and Detector Concepts for the Electron-Ion Collider}: {EIC Yellow Report}",
    eprint = "2103.05419",
    archivePrefix = "arXiv",
    primaryClass = "physics.ins-det",
    reportNumber = "BNL-220990-2021-FORE, JLAB-PHY-21-3198, LA-UR-21-20953",
    doi = "10.1016/j.nuclphysa.2022.122447",
    journal = "Nucl. Phys. A",
    volume = "1026",
    pages = "122447",
    year = "2022"
}

@article{Bertulani:2005ru,
    author = "Bertulani, Carlos A. and Klein, Spencer R. and Nystrand, Joakim",
    title = "{Physics of ultra-peripheral nuclear collisions}",
    eprint = "nucl-ex/0502005",
    archivePrefix = "arXiv",
    doi = "10.1146/annurev.nucl.55.090704.151526",
    journal = "Ann. Rev. Nucl. Part. Sci.",
    volume = "55",
    pages = "271--310",
    year = "2005"
}

@misc{OSPool,
  doi = {10.21231/906P-4D78},
  url = {https://osg-htc.org/services/open_science_pool.html},
  author = {{OSG}},
  title = {OSPool},
  publisher = {OSG},
  year = {2006}
}

@misc{OSDF,
  doi = {10.21231/0KVZ-VE57},
  url = {https://osdf.osg-htc.org/},
  author = {{OSG}},
  title = {Open Science Data Federation},
  publisher = {OSG},
  year = {2015}
}

@article{Mantysaari:2026dps,
    author = {M{\"a}ntysaari, Heikki and Roch, Hendrik and Schenke, Bj{\"o}rn and Shen, Chun and Zhao, Wenbin},
    title = "{Nuclear structure and saturation effects from diffractive vector meson production}",
    eprint = "2605.00454",
    archivePrefix = "arXiv",
    primaryClass = "hep-ph",
    month = "5",
    year = "2026"
}

@article{Dyndal:2026uvm,
    author = "Dyndal, M. and Harland-Lang, L. A.",
    title = "{A First Account of the Impact of Ion Electromagnetic Dissociation on Event Exclusivity in Ultraperipheral LHC Collisions}",
    eprint = "2604.19879",
    archivePrefix = "arXiv",
    primaryClass = "hep-ph",
    month = "4",
    year = "2026"
}

@article{Mantysaari:2025ltq,
    author = {M{\"a}ntysaari, Heikki and Roch, Hendrik and Salazar, Farid and Schenke, Bj{\"o}rn and Shen, Chun and Zhao, Wenbin},
    title = "{Global Bayesian analysis of J/{\ensuremath{\psi}} photoproduction on proton and lead targets}",
    eprint = "2507.14087",
    archivePrefix = "arXiv",
    primaryClass = "hep-ph",
    doi = "10.1103/pcmz-dyz1",
    journal = "Phys. Rev. D",
    volume = "113",
    number = "1",
    pages = "014038",
    year = "2026"
}

@article{Gribov:1983ivg,
    author = "Gribov, L. V. and Levin, E. M. and Ryskin, M. G.",
    title = "{Semihard Processes in QCD}",
    doi = "10.1016/0370-1573(83)90022-4",
    journal = "Phys. Rept.",
    volume = "100",
    pages = "1--150",
    year = "1983"
}

@article{Mueller:1985wy,
    author = "Mueller, Alfred H. and Qiu, Jian-wei",
    title = "{Gluon Recombination and Shadowing at Small Values of x}",
    reportNumber = "CU-TP-322",
    doi = "10.1016/0550-3213(86)90164-1",
    journal = "Nucl. Phys. B",
    volume = "268",
    pages = "427--452",
    year = "1986"
}

@article{McLerran:1993ni,
    author = "McLerran, Larry D. and Venugopalan, Raju",
    title = "{Computing quark and gluon distribution functions for very large nuclei}",
    eprint = "hep-ph/9309289",
    archivePrefix = "arXiv",
    reportNumber = "TPI-MINN-93-44-T, NUC-MINN-93-24-T, HEP-UMN-TH-1220-93",
    doi = "10.1103/PhysRevD.49.2233",
    journal = "Phys. Rev. D",
    volume = "49",
    pages = "2233--2241",
    year = "1994"
}

@article{McLerran:1993ka,
    author = "McLerran, Larry D. and Venugopalan, Raju",
    title = "{Gluon distribution functions for very large nuclei at small transverse momentum}",
    eprint = "hep-ph/9311205",
    archivePrefix = "arXiv",
    reportNumber = "TPI-MINN-93-52-T, NUC-MINN-93-28-T, UMN-TH-1224-93",
    doi = "10.1103/PhysRevD.49.3352",
    journal = "Phys. Rev. D",
    volume = "49",
    pages = "3352--3355",
    year = "1994"
}

@inbook{Iancu:2003xm,
    author = "Iancu, Edmond and Venugopalan, Raju",
    editor = "Hwa, Rudolph C. and Wang, Xin-Nian",
    title = "{The Color glass condensate and high-energy scattering in QCD}",
    booktitle = "{Quark-gluon plasma 4}",
    eprint = "hep-ph/0303204",
    archivePrefix = "arXiv",
    doi = "10.1142/9789812795533_0005",
    pages = "249--3363",
    month = "3",
    year = "2003"
}

@article{Morreale:2021pnn,
    author = "Morreale, Astrid and Salazar, Farid",
    title = "{Mining for Gluon Saturation at Colliders}",
    eprint = "2108.08254",
    archivePrefix = "arXiv",
    primaryClass = "hep-ph",
    reportNumber = "Universe 2021",
    doi = "10.3390/universe7080312",
    journal = "Universe",
    volume = "7",
    number = "8",
    pages = "312",
    year = "2021"
}

@article{Garcia-Montero:2025hys,
    author = {Garcia-Montero, Oscar and Schlichting, S{\"o}ren},
    title = "{Effective theories for nuclei at high energies}",
    eprint = "2502.09721",
    archivePrefix = "arXiv",
    primaryClass = "hep-ph",
    doi = "10.1140/epja/s10050-025-01523-7",
    journal = "Eur. Phys. J. A",
    volume = "61",
    number = "3",
    pages = "54",
    year = "2025"
}

@article{Ryskin:1992ui,
    author = "Ryskin, M. G.",
    title = "{Diffractive $\mathrm{J}/\psi$ electroproduction in LLA QCD}",
    reportNumber = "LU-TP-92-12",
    doi = "10.1007/BF01555742",
    journal = "Z. Phys. C",
    volume = "57",
    pages = "89--92",
    year = "1993"
}

@article{Mantysaari:2020axf,
    author = {M{\"a}ntysaari, Heikki},
    title = "{Review of proton and nuclear shape fluctuations at high energy}",
    eprint = "2001.10705",
    archivePrefix = "arXiv",
    primaryClass = "hep-ph",
    doi = "10.1088/1361-6633/aba347",
    journal = "Rept. Prog. Phys.",
    volume = "83",
    number = "8",
    pages = "082201",
    year = "2020"
}

@article{Mantysaari:2017cni,
    author = {M{\"a}ntysaari, Heikki and Schenke, Bj{\"o}rn and Shen, Chun and Tribedy, Prithwish},
    title = "{Imprints of fluctuating proton shapes on flow in proton-lead collisions at the LHC}",
    eprint = "1705.03177",
    archivePrefix = "arXiv",
    primaryClass = "nucl-th",
    doi = "10.1016/j.physletb.2017.07.038",
    journal = "Phys. Lett. B",
    volume = "772",
    pages = "681--686",
    year = "2017"
}

@article{Mantysaari:2018zdd,
    author = {M{\"a}ntysaari, Heikki and Schenke, Bj{\"o}rn},
    title = "{Confronting impact parameter dependent JIMWLK evolution with HERA data}",
    eprint = "1806.06783",
    archivePrefix = "arXiv",
    primaryClass = "hep-ph",
    doi = "10.1103/PhysRevD.98.034013",
    journal = "Phys. Rev. D",
    volume = "98",
    number = "3",
    pages = "034013",
    year = "2018"
}

@article{Mantysaari:2023qsq,
    author = {M{\"a}ntysaari, Heikki and Schenke, Bj{\"o}rn and Shen, Chun and Zhao, Wenbin},
    title = "{Multiscale Imaging of Nuclear Deformation at the Electron-Ion Collider}",
    eprint = "2303.04866",
    archivePrefix = "arXiv",
    primaryClass = "nucl-th",
    doi = "10.1103/PhysRevLett.131.062301",
    journal = "Phys. Rev. Lett.",
    volume = "131",
    number = "6",
    pages = "062301",
    year = "2023"
}

@article{Lappi:2013am,
    author = "Lappi, T. and Mantysaari, H.",
    title = "{$\mathrm{J}/\psi$ production in ultraperipheral Pb+Pb and $p$+Pb collisions at energies available at the CERN Large Hadron Collider}",
    eprint = "1301.4095",
    archivePrefix = "arXiv",
    primaryClass = "hep-ph",
    doi = "10.1103/PhysRevC.87.032201",
    journal = "Phys. Rev. C",
    volume = "87",
    number = "3",
    pages = "032201",
    year = "2013"
}

@article{Mantysaari:2022sux,
    author = {M{\"a}ntysaari, Heikki and Salazar, Farid and Schenke, Bj{\"o}rn},
    title = "{Nuclear geometry at high energy from exclusive vector meson production}",
    eprint = "2207.03712",
    archivePrefix = "arXiv",
    primaryClass = "hep-ph",
    doi = "10.1103/PhysRevD.106.074019",
    journal = "Phys. Rev. D",
    volume = "106",
    number = "7",
    pages = "074019",
    year = "2022"
}

@article{Cepila:2025rkn,
    author = "Cepila, J. and Contreras, J. G. and Vaculciak, M.",
    title = "{Exclusive quarkonium photoproduction: Predictions with the Balitsky-Kovchegov equation including the full impact-parameter dependence}",
    eprint = "2501.09462",
    archivePrefix = "arXiv",
    primaryClass = "hep-ph",
    doi = "10.1103/PhysRevD.111.056002",
    journal = "Phys. Rev. D",
    volume = "111",
    number = "5",
    pages = "056002",
    year = "2025"
}

@article{Mantysaari:2023xcu,
    author = {M{\"a}ntysaari, Heikki and Salazar, Farid and Schenke, Bj{\"o}rn},
    title = "{Energy dependent nuclear suppression from gluon saturation in exclusive vector meson production}",
    eprint = "2312.04194",
    archivePrefix = "arXiv",
    primaryClass = "hep-ph",
    reportNumber = "INT-PUB-24-008",
    doi = "10.1103/PhysRevD.109.L071504",
    journal = "Phys. Rev. D",
    volume = "109",
    number = "7",
    pages = "L071504",
    year = "2024"
}

@article{Mantysaari:2024zxq,
    author = {M{\"a}ntysaari, Heikki and Penttala, Jani and Salazar, Farid and Schenke, Bj{\"o}rn},
    title = "{Finite-size effects on small-x evolution and saturation in proton and nuclear targets}",
    eprint = "2411.13533",
    archivePrefix = "arXiv",
    primaryClass = "hep-ph",
    doi = "10.1103/PhysRevD.111.054033",
    journal = "Phys. Rev. D",
    volume = "111",
    number = "5",
    pages = "054033",
    year = "2025"
}

@article{Kowalski:2006hc,
    author = "Kowalski, H. and Motyka, L. and Watt, G.",
    title = "{Exclusive diffractive processes at HERA within the dipole picture}",
    eprint = "hep-ph/0606272",
    archivePrefix = "arXiv",
    reportNumber = "DESY-06-095",
    doi = "10.1103/PhysRevD.74.074016",
    journal = "Phys. Rev. D",
    volume = "74",
    pages = "074016",
    year = "2006"
}

@article{Mueller:2001uk,
    author = "Mueller, Alfred H.",
    title = "{A Simple derivation of the JIMWLK equation}",
    eprint = "hep-ph/0110169",
    archivePrefix = "arXiv",
    reportNumber = "CU-TP-1031",
    doi = "10.1016/S0370-2693(01)01343-0",
    journal = "Phys. Lett. B",
    volume = "523",
    pages = "243--248",
    year = "2001"
}

@article{Lappi:2012vw,
    author = {Lappi, T. and M{\"a}ntysaari, H.},
    title = "{On the running coupling in the JIMWLK equation}",
    eprint = "1212.4825",
    archivePrefix = "arXiv",
    primaryClass = "hep-ph",
    doi = "10.1140/epjc/s10052-013-2307-z",
    journal = "Eur. Phys. J. C",
    volume = "73",
    number = "2",
    pages = "2307",
    year = "2013"
}

@article{Karamanis:2022alw,
    author = "Karamanis, Minas and Beutler, Florian and Peacock, John A. and Nabergoj, David and Seljak, Uros",
    title = "{Accelerating astronomical and cosmological inference with preconditioned Monte Carlo}",
    eprint = "2207.05652",
    archivePrefix = "arXiv",
    primaryClass = "astro-ph.IM",
    doi = "10.1093/mnras/stac2272",
    journal = "Mon. Not. Roy. Astron. Soc.",
    volume = "516",
    number = "2",
    pages = "1644--1653",
    year = "2022"
}

@article{Karamanis:2022ksp,
    author = "Karamanis, Minas and Nabergoj, David and Beutler, Florian and Peacock, John A. and Seljak, Uros",
    title = "{pocoMC: A Python package for accelerated Bayesian inference in astronomy and cosmology}",
    eprint = "2207.05660",
    archivePrefix = "arXiv",
    primaryClass = "astro-ph.IM",
    doi = "10.21105/joss.04634",
    journal = "J. Open Source Softw.",
    volume = "7",
    number = "79",
    pages = "4634",
    year = "2022"
}

@article{ALICE:2014eof,
    author = "Abelev, Betty Bezverkhny and others",
    collaboration = "ALICE",
    title = "{Exclusive $\mathrm{J/}\psi$ photoproduction off protons in ultra-peripheral p-Pb collisions at $\sqrt{s_{\rm NN}}=5.02$ TeV}",
    eprint = "1406.7819",
    archivePrefix = "arXiv",
    primaryClass = "nucl-ex",
    reportNumber = "CERN-PH-EP-2014-149",
    doi = "10.1103/PhysRevLett.113.232504",
    journal = "Phys. Rev. Lett.",
    volume = "113",
    number = "23",
    pages = "232504",
    year = "2014"
}

@article{ALICE:2018oyo,
    author = "Acharya, Shreyasi and others",
    collaboration = "ALICE",
    title = "{Energy dependence of exclusive $\mathrm {J}/\psi $ photoproduction off protons in ultra-peripheral p{\textendash}Pb collisions at $\sqrt{s_{\mathrm {\scriptscriptstyle NN}}} = 5.02$ TeV}",
    eprint = "1809.03235",
    archivePrefix = "arXiv",
    primaryClass = "nucl-ex",
    reportNumber = "CERN-EP-2018-236",
    doi = "10.1140/epjc/s10052-019-6816-2",
    journal = "Eur. Phys. J. C",
    volume = "79",
    number = "5",
    pages = "402",
    year = "2019"
}

@article{H1:2005dtp,
    author = "Aktas, A. and others",
    collaboration = "H1",
    title = "{Elastic J/psi production at HERA}",
    eprint = "hep-ex/0510016",
    archivePrefix = "arXiv",
    reportNumber = "DESY-05-161",
    doi = "10.1140/epjc/s2006-02519-5",
    journal = "Eur. Phys. J. C",
    volume = "46",
    pages = "585--603",
    year = "2006"
}

@article{H1:2013okq,
    author = "Alexa, C. and others",
    collaboration = "H1",
    title = "{Elastic and Proton-Dissociative Photoproduction of $\mathrm{J}/\psi$ Mesons at HERA}",
    eprint = "1304.5162",
    archivePrefix = "arXiv",
    primaryClass = "hep-ex",
    reportNumber = "DESY-13-058",
    doi = "10.1140/epjc/s10052-013-2466-y",
    journal = "Eur. Phys. J. C",
    volume = "73",
    number = "6",
    pages = "2466",
    year = "2013"
}

@article{ALICE:2023jgu,
    author = "Acharya, Shreyasi and others",
    collaboration = "ALICE",
    title = "{Energy dependence of coherent photonuclear production of J/{\ensuremath{\psi}} mesons in ultra-peripheral Pb-Pb collisions at $ \sqrt{{\textrm{s}}_{\textrm{NN}}} $ = 5.02 TeV}",
    eprint = "2305.19060",
    archivePrefix = "arXiv",
    primaryClass = "nucl-ex",
    reportNumber = "CERN-EP-2023-100",
    doi = "10.1007/JHEP10(2023)119",
    journal = "JHEP",
    volume = "10",
    pages = "119",
    year = "2023"
}

@article{CMS:2023snh,
    author = "Tumasyan, Armen and others",
    collaboration = "CMS",
    title = "{Probing Small Bjorken-x Nuclear Gluonic Structure via Coherent J/{\ensuremath{\psi}} Photoproduction in Ultraperipheral Pb-Pb Collisions at $\sqrt{s_{NN}}=5.02${\,}{\,}TeV}",
    eprint = "2303.16984",
    archivePrefix = "arXiv",
    primaryClass = "nucl-ex",
    reportNumber = "CMS-HIN-22-002, CERN-EP-2023-031",
    doi = "10.1103/PhysRevLett.131.262301",
    journal = "Phys. Rev. Lett.",
    volume = "131",
    number = "26",
    pages = "262301",
    year = "2023"
}

@article{ALICE:2021tyx,
    author = "Acharya, Shreyasi and others",
    collaboration = "ALICE",
    title = "{First measurement of the |$t$|-dependence of coherent $J/\psi$ photonuclear production}",
    eprint = "2101.04623",
    archivePrefix = "arXiv",
    primaryClass = "nucl-ex",
    reportNumber = "CERN-EP-2021-003",
    doi = "10.1016/j.physletb.2021.136280",
    journal = "Phys. Lett. B",
    volume = "817",
    pages = "136280",
    year = "2021"
}

@article{ALICE:2023gcs,
    author = "Acharya, Shreyasi and others",
    collaboration = "ALICE",
    title = "{First Measurement of the |t| Dependence of Incoherent J/{\ensuremath{\psi}} Photonuclear Production}",
    eprint = "2305.06169",
    archivePrefix = "arXiv",
    primaryClass = "nucl-ex",
    reportNumber = "CERN-EP-2023-080",
    doi = "10.1103/PhysRevLett.132.162302",
    journal = "Phys. Rev. Lett.",
    volume = "132",
    number = "16",
    pages = "162302",
    year = "2024"
}

@article{ZEUS:2002wfj,
    author = "Chekanov, S. and others",
    collaboration = "ZEUS",
    title = "{Exclusive photoproduction of $\mathrm{J}/\psi$ mesons at HERA}",
    eprint = "hep-ex/0201043",
    archivePrefix = "arXiv",
    reportNumber = "DESY-02-008",
    doi = "10.1007/s10052-002-0953-7",
    journal = "Eur. Phys. J. C",
    volume = "24",
    pages = "345--360",
    year = "2002"
}

@article{LHCb:2018rcm,
    author = "Aaij, Roel and others",
    collaboration = "LHCb",
    title = "{Central exclusive production of $J/\psi$ and $\psi(2S)$ mesons in $pp$ collisions at $\sqrt{s}=13~$TeV}",
    eprint = "1806.04079",
    archivePrefix = "arXiv",
    primaryClass = "hep-ex",
    reportNumber = "LHCB-PAPER-2018-011, LHCb-PAPER-2018-011, CERN-EP-2018-152",
    doi = "10.1007/JHEP10(2018)167",
    journal = "JHEP",
    volume = "10",
    pages = "167",
    year = "2018"
}

@article{Gordon:2007xm,
    author = "Gordon, Christopher and Trotta, Roberto",
    title = "{Bayesian Calibrated Significance Levels Applied to the Spectral Tilt and Hemispherical Asymmetry}",
    eprint = "0706.3014",
    archivePrefix = "arXiv",
    primaryClass = "astro-ph",
    doi = "10.1111/j.1365-2966.2007.12707.x",
    journal = "Mon. Not. Roy. Astron. Soc.",
    volume = "382",
    pages = "1859--1863",
    year = "2007"
}

@article{Trotta:2008qt,
    author = "Trotta, Roberto",
    title = "{Bayes in the sky: Bayesian inference and model selection in cosmology}",
    eprint = "0803.4089",
    archivePrefix = "arXiv",
    primaryClass = "astro-ph",
    doi = "10.1080/00107510802066753",
    journal = "Contemp. Phys.",
    volume = "49",
    pages = "71--104",
    year = "2008"
}

@misc{ipglasma_jimwlk_code,
  author={Mäntysaari, Heikki and Schenke, Björn and Shen, Chun and Zhao, Wenbin},
  title        = {{JIMWLK + IP-Glasma}},
  year         = {2025},
  howpublished = {\url{https://github.com/schenke/ipglasma/tree/ipglasma_jimwlk}},
  note         = {GitHub repository},
}

@misc{subnucleondiffraction_code,
  author={Mäntysaari, Heikki},
  title        = {{Subnucleondiffraction}},
  year         = {2026},
  howpublished = {\url{https://github.com/hejajama/subnucleondiffraction}},
  note         = {GitHub repository},
}

@misc{IPGlasmaFramework,
  author       = {Shen, Chun},
  title        = {{IPGlasmaFramework}},
  howpublished = {\url{https://github.com/chunshen1987/IPGlasmaFramework}},
  note         = {GitHub repository},
  year         = {2026},
  urldate      = {2026-06-12}
}

@dataset{roch_2026_20696856,
  author       = {Roch, Hendrik and
                  Mäntysaari, Heikki and
                  Schenke, Björn and
                  Shen, Chun and
                  Zhao, Wenbin},
  title        = {Revisiting the role of saturation in diffractive
                   vector meson production
                  },
  month        = jun,
  year         = 2026,
  publisher    = {Zenodo},
  doi          = {10.5281/zenodo.20696856},
  url          = {https://doi.org/10.5281/zenodo.20696856},
}

@article{Mantysaari:2025ujz,
    author = {M{\"a}ntysaari, Heikki and Roch, Hendrik and Salazar, Farid and Schenke, Bj{\"o}rn and Shen, Chun and Zhao, Wenbin},
    title = "{Nuclear suppression in diffractive vector meson production within the color glass condensate framework}",
    eprint = "2508.21562",
    archivePrefix = "arXiv",
    primaryClass = "hep-ph",
    doi = "10.1051/epjconf/202636413007",
    journal = "EPJ Web Conf.",
    volume = "364",
    pages = "13007",
    year = "2026"
}

@inproceedings{Mantysaari:2025cok,
    author = {M{\"a}ntysaari, Heikki and Roch, Hendrik and Salazar, Farid and Schenke, Bj{\"o}rn and Shen, Chun and Zhao, Wenbin},
    title = "{Nuclear Suppression in Diffractive Vector Meson Production within the Color Glass Condensate Framework}",
    booktitle = "{2nd International Workshop on the Physics of Ultra Peripheral Collisions}",
    eprint = "2509.13015",
    archivePrefix = "arXiv",
    primaryClass = "hep-ph",
    month = "9",
    year = "2025"
}

\end{document}